\documentclass[]{spie}  

\usepackage{amsmath,amsfonts,amssymb}
\usepackage{graphicx}
\usepackage[colorlinks=true, allcolors=blue]{hyperref}
\usepackage{amsmath,amsfonts,amssymb}
\usepackage{multicol}
\usepackage{tabu}
\usepackage{float}
\usepackage{textcomp,gensymb}

\title{Program objectives and specifications
for the Ultra-Fast Astronomy observatory}

\author[a]{Siyang Li}
\author[a-h]{George F. Smoot}
\author[b,d]{Bruce Grossan}
\author[e]{Albert Wai Kit Lau}
\author[d]{Marzhan Bekbalanova}
\author[d]{Mehdi Shafiee}
\author[c]{Thorsten Stezelberger}
\affil[a]{Department of Physics, University of California, Berkeley, USA}
\affil[b]{Space Sciences Laboratory, University of California, Berkeley, USA}
\affil[c]{Lawrence Berkeley National Laboratory, USA}
\affil[d]{Energetic Cosmos Laboratory, Nazarbayev University, Kazakhstan}
\affil[e]{Department of Physics, Hong Kong University of Science and Technology, China}
\affil[f]{Institute for Advanced Study, Hong Kong University of Science and Technology, China}
\affil[g]{Department of Physics, Universit\'e Paris Diderot, France}
\affil[h]{Paris Centre for Cosmological Physics, Universit\'e Paris, France}

\authorinfo{Further author information: (Send correspondence to S.L.)\\S.L.: E-mail: seanli@berkeley.edu}

\begin{document} 
\maketitle

\begin{abstract}

We present program objectives and specifications for the first generation Ultra-Fast Astronomy (UFA) observatory which will explore a new astrophysical phase space by characterizing the variability of the optical (320 nm - 650 nm) sky in the millisecond to nanosecond timescales. One of the first objectives of the UFA observatory will be to search for optical counterparts to fast radio bursts (FRB) that can be used to identify the origins of FRB and probe the epoch of reionization and baryonic matter in the interstellar and intergalactic mediums. The UFA camera will consist of two single-photon resolution fast-response detector 16x16 arrays operated in coincidence mounted on the 0.7 meter Nazarbayev University Transient Telescope at the Assy-Turgen Astrophysical Observatory (NUTTelA-TAO) located near Almaty, Kazakhstan. We are currently developing two readout systems that can measure down to the microsecond and nanosecond timescales and characterizing two silicon photomultipliers (SiPM) and one photomultiplier tube (PMT) to compare the detectors for the UFA observatory and astrophysical observations in general.

\end{abstract}

\section{INTRODUCTION}
\label{sec:intro}  

Historically, searching in previously unexplored phase spaces has led to discoveries of new astrophysical phenomena. For instance, the discovery of fast radio bursts (FRB) occurred after archival data from the Parkes radio telescope was searched in the millisecond timescale\cite{Lorimer777}. While undiscovered sub-second optical phenomena may exist, few measurements have been conducted in the optical sub-second regime\cite{Horowitz2001, Eikenberry_1997, Leung:2018} due to the long read times and noise penalties of most standard array imagers such as charge coupled devices (CCD).

One example of a possible undiscovered sub-second optical transient is an optical counterpart to FRB. FRB are high energy millisecond duration radio transients of unknown astrophysical origin. To date, 82 FRB\cite{Petroff2016} with frequencies ranging from 400 MHz\cite{Amiri2019} to 8 GHz\cite{Gajjar2018} have been discovered using large dish radio telescopes. Several theories on the origins of FRB have been proposed including white dwarf mergers \cite{Kashiyama_2013}, neutron star collisions \cite{Geng_2015}, supergiant pulses from neutron stars \cite{10.1093/mnras/stv2948}, giant magnetar flares \cite{Thornton2013,Pen_2015}, neutron stars collapsing into black holes \cite{Falcke2013}, young pulsars in supernova remnants \cite{Conner2016, Piro:2016aac}, and coherent curvature radiation emitted from magnetars \cite{Kumar2017}. The discovery of repeating FRB\cite{Spitler2016, Amiri2019b} indicate that FRB, or at least a subclass of FRB, originate from non-cataclysmic origins. An observable optical counterpart would allow us to investigate and potentially identify the origins and emission mechanisms of FRB. In addition, as FRB exhibit high dispersion measures that point to an extragalactic origin, optical counterparts to FRB would also allow us to probe the epoch of reionization and baryonic matter in the interstellar and intergalactic mediums.


To date, searches for optical counterparts to FRB have been conducted down to the millisecond timescale. Hardy et al. searched for optical counterparts to FRB in the direction of the repeating FRB 121102 in coordination with the Arecibo Observatory and found no optical counterparts in their 70.7 millisecond timescale\cite{Hardy2017}. However, searches for optical counterpart emission would require much shorter integration times than in radio due to dispersion and pulse-broadening by electrons in the line of sight to the source. Searches for sub-millisecond duration flashes using millisecond duration cameras would be severely limited by the signal to noise ratio and would require the hardware needed to explore shorter timescales.


In this paper, we describe program objectives and specifications of the first generation Ultra-Fast Astronomy (UFA) observatory which will utilize single-photon counting detectors to characterize the sky and search for optical counterparts to FRB in the sub-second time domain. 


\section{PROGRAM OBJECTIVES}

The UFA observatory will explore a new astrophysical phase space by characterizing the optical (320 nm - 650 nm) sky in the millisecond to nanosecond timescales using single-photon counting technology. Exploring the universe down to the nanosecond timescale will allow us to search for undiscovered sub-second astrophysical phenomena such as surface convection on white dwarfs, variability close to black holes, and interstellar laser communications from intelligent extraterrestrial civilizations.

One of our first objectives will be to search for and further constrain the limits of optical counterparts to FRB. We will search for millisecond to nanosecond optical signals in the direction of repeating FRBs. An observable optical counterpart would allow us to potentially identify the origins of FRB and probe the baryonic matter in the intergalactic medium and Epoch of Reionization. The UFA program will improve upon previous searches for optical counterparts to FRB by searching in a faster millisecond to nanosecond timescale with higher sensitivity.

\section{INSTRUMENT}

\subsection{Telescope}

The Nazarbayev University Transient Telescope at the Assy-Turgen Astrophysical Observatory (NUTTelA-TAO) is a Corrected Dall-Kirkham telescope from Planewave Instruments (CDK700) having an aperture of 700 mm and a focal ratio of 6.5. Installation of the NUTTelA-TAO at the Assy-Turgan Observatory completed in October 2018. The UFA instrument will share the NUTTelA-TAO with the Burst Simultaneous Three-channel Instrument (BSTI) camera which will use electron multiplied charge coupled devices (EMCCD) to measure optical prompt emission spectra from gamma ray bursts \cite{Grossan2017, Grossan2019}. A plane mirror will redirect light to the UFA camera when the BSTI is not in use.


\subsection{Camera}

\begin{figure}
\begin{center}
\includegraphics[height=9cm]{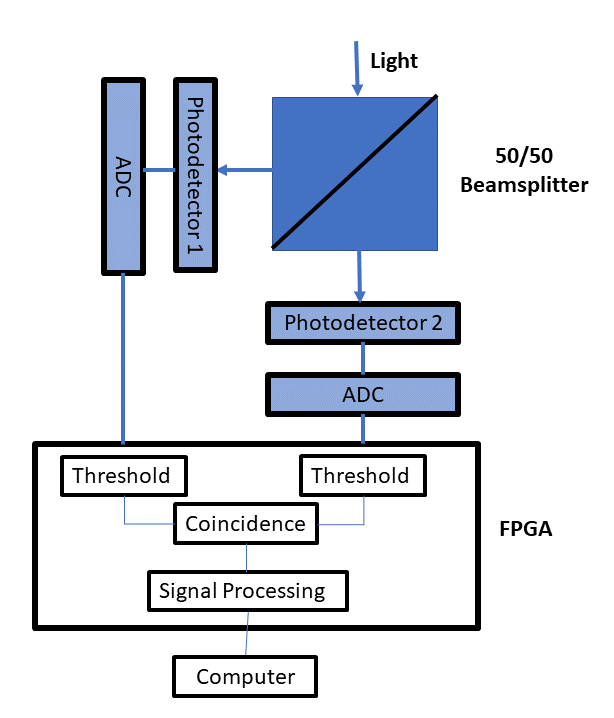}
\end{center}
\caption[example] 
{Coincidence scheme that will be used for the UFA camera. Two single-photon counting detectors will be placed at a 90\degree angle with a 50/50 non-polarizing beamsplitter. Events will be processed using coincidence logic.}
\label{fig:Coincidence}
\end{figure}

The UFA camera will consist of two single-photon resolution fast-response detector 16x16 arrays operated in coincidence using a 50/50 non-polarizing beamsplitter (Fig. \ref{fig:Coincidence}). Only events occurring simultaneously within a coincidence window in both detectors will be recorded. This setup will help reduce the number of false alarms caused by both random sky background and internal detector noise such as dark current, crosstalk, and afterpulsing.

Photomultiplier tubes (PMT) and silicon photomultipliers (SiPM) are both particularly advantageous for the UFA camera due to their nanosecond duration timing resolutions. In comparison, CCD are limited by their read noise and the fastest EMCCDs currently are on the order of 0.1 s. In addition, both PMT and SiPM are capable of detecting faint single-photon events with a sky background. As pulses produced by these detectors have heights proportional to the number of detected photons high-flux photon events can be distinguished from random single or low-flux photon events by placing a threshold.

The overall advantage of using either PMT or SiPM over the other for the UFA camera is not yet clear. Compared to PMT, SiPM have comparable gains and higher photon detection efficiencies, affordability, and durability. However, SiPMs also have higher dark count rates. We will first characterize each detector type before selecting an optimal detector for the UFA observatory. Although only one detector will be used for the UFA observatory, this study will also provide a reference for future observatories deciding between PMT and SiPM and searching for fast astrophysical transients. 

\begin{table}[H]
\begin{center}
\begin{tabu}  { | X[c] | X[c] | X[c] |  X[c] |  }
 \hline
 \textbf{Parameter} & \textbf{H9500 PMT} & \textbf{S13360 MPPC}& \textbf{S14161 MPPC}  \\
  \hline
 Spectral Range & 300 nm - 650 nm  & 320 nm - 900 nm  & 270 nm - 900 nm\\
 \hline
Recommended Operating Voltage  & 1000V & 53V + 3V & 38V + 2.7V\\

  \hline
 Channel Size & 3.04 mm x 3.04 mm & 3 mm x 3 mm & 3 mm x 3mm\\
   \hline
Gain & 1.5x10\textsuperscript{6} & 1.7x10\textsuperscript{6} & 2.5x10\textsuperscript{6} \\
 \hline
\end{tabu}
\end{center}
\caption{Typical detector parameters of the H9500 PMT, S13360 series MPPC, and S14161 series MPPC from Hamamatsu Photonics K. K.\cite{datasheet,datasheetb,datasheetPMT}. The recommended operating voltage for MPPC is the breakdown voltage plus the recommended over voltage. Gain is given at the recommended operating voltage.}
\label{tab:parameters}
\end{table}

\subsubsection{Photomultiplier Tube}
\label{sec:PMT}

PMT are vacuum tube photodetectors with single-photon counting capabilities that have traditionally been used for a variety of low-light applications in fields such particle physics, spectroscopy, microscopy, medical imaging, and astronomy. In astronomy, PMT have been used to search for astrophysical events such as cosmic ray showers \cite{Adams2015}, gamma ray bursts \cite{Atwood:2009ez}, and extraterrestrial technosignatures \cite{HOWARD200778}.

\begin{figure} [H]
\begin{center}
\includegraphics[height=8cm]{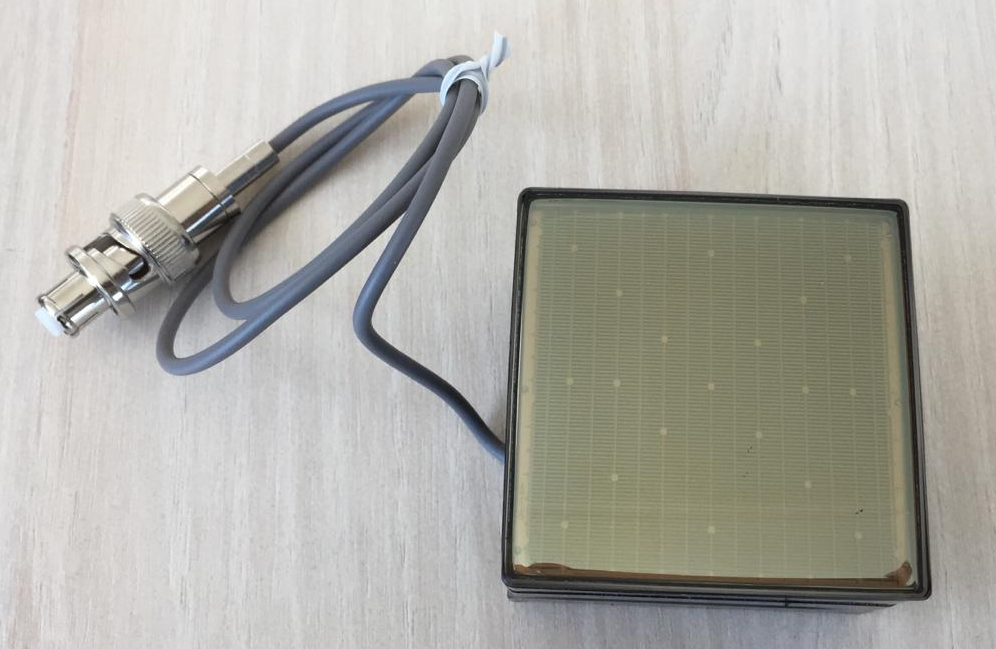}
\end{center}
\caption[example] 
{H9500 PMT from Hamamatsu Photonics K. K.. This PMT will be characterized at Nazarbayev University.}
\label{fig:PMT}
\end{figure}

Our candidate PMT is the H9500 multianode PMT 16x16 array from Hamamatsu Photonics K.K. (Fig. \ref{fig:PMT}). Detector parameters can be seen in Table \ref{tab:parameters}. We selected this PMT due to its large number of channels, high quantum efficiency (24$\%$ at 420 nm)\cite{datasheetPMT} compared to similar PMTs, and convenient metal packaging configuration. To compare this PMT with our candidate SiPMs (section \ref{sec:SiPM}), we will characterize its breakdown voltage, gain, dark count rate, crosstalk probability, afterpulsing, photon detection efficiency, linearity, saturation, and photon detection efficiency inside a dark box test stand at Nazarbayev University. 

\subsubsection{Silicon Photomultiplier}
\label{sec:SiPM}

The SiPM is a p-n junction solid state photodetector with single-photon counting capabilities consisting of single photon avalanche photodiodes (SPAD) connected in parallel. SiPM and SPAD are beginning to replace PMT in searches for astrophysical transients such as cosmic ray showers \cite{Teshima2007,Guberman2017} and extraterrestrial technosignatures \cite{Wright2018,Li2019}. 

\begin{figure} [H]
\begin{center}
\begin{multicols}{2}
\includegraphics[height=6cm]{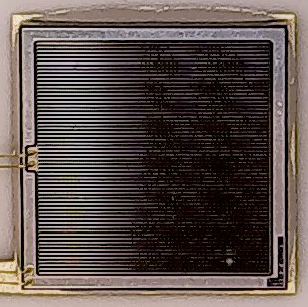}\par 
\includegraphics[height=6cm]{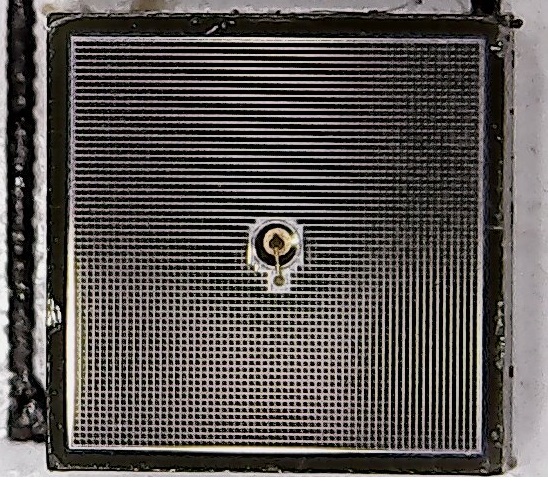}\par 
\end{multicols}
\end{center}
\caption{Both MPPC shown have 3mm x 3mm active areas and 50 $\mu$m pixel pitches. LEFT: S13360-3050CS MPPC RIGHT: S14160-3050HS MPPC.}
\label{fig:detectors}
\end{figure}

We have narrowed down our candidate SiPMs to two SiPM manufactured by Hamamatsu Photonics K. K.: the S13360 series Multi-Pixel Photon Counter (MPPC) and S14161 series MPPC. We selected these SiPMs due to their high photon detection efficiencies compared to SiPMs of similar microcell sizes and active areas from other companies. While the S14161 MPPC has a higher reported peak photon detection efficiency than the S13360 MPPC (50$\%$ at 2.7V over voltage vs. 40$\%$ at 3V over voltage, both at 450nm), it also has a higher reported dark count rate (1.1 Mcps at 2.7V over voltage vs. 500 kcps at 3V over voltage)\cite{datasheet,datasheetb}. To compare these two detectors in further detail, we will characterize the dark count rate, crosstalk probability, photon detection efficiency, linear range, and saturation of the single channel versions of the S13360 series MPPC (S13360-3050CS)\cite{Li2019b} and the S14160 series MPPC (S14160-3050HS) at the Hong Kong University of Science and Technology. Images of the S13360-3050CS MPPC and S14160-3050HS MPPC can be seen in Fig. \ref{fig:detectors}. Because the largest arrays for the S13360 and S14160 series offered by Hamamatsu Photonics K. K. are limited to 8x8, we will tile four 8x8 arrays to create a 16x16 array for the UFA camera.

\subsection{Readout}

\begin{figure} [H]
\begin{center}
\includegraphics[height=8cm]{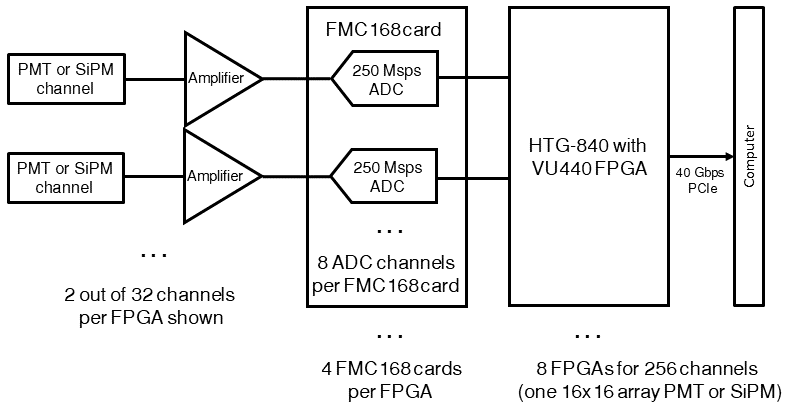}
\end{center}
\caption[example] 
{Schematic of the readout system using the HTG-840 development platform from Xilinx and FMC168 Mezzanine Card from Abaco Systems.}
\label{fig:HTG840}
\end{figure}

\begin{figure} [H]
\begin{center}
\begin{multicols}{2}
\includegraphics[height=6cm]{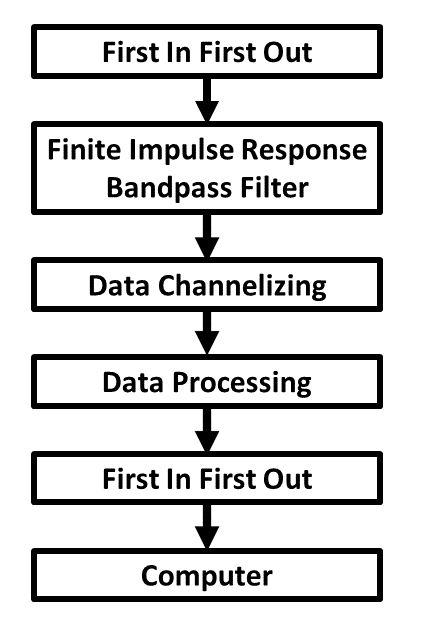}\par 
\includegraphics[height=6cm]{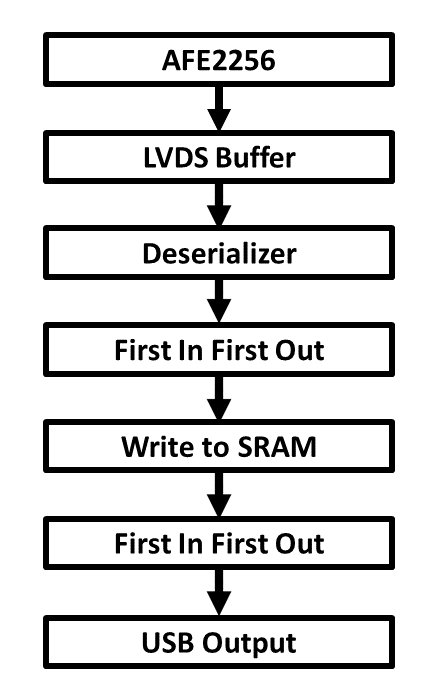}\par 
\end{multicols}
\end{center}
\caption[example] 
{LEFT: Data processing flow that will be used with the HTG-840 and FMC168 readout system. RIGHT: Data processing flow that will be used with the AFE2256EVM readout system.}
\label{fig:DataProcess}
\end{figure}

\begin{figure} [H]
\begin{center}
\begin{multicols}{2}
\includegraphics[height=6cm]{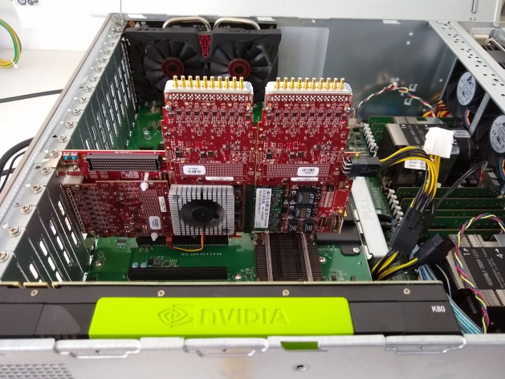}\par 
\includegraphics[angle= 90, height=6cm]{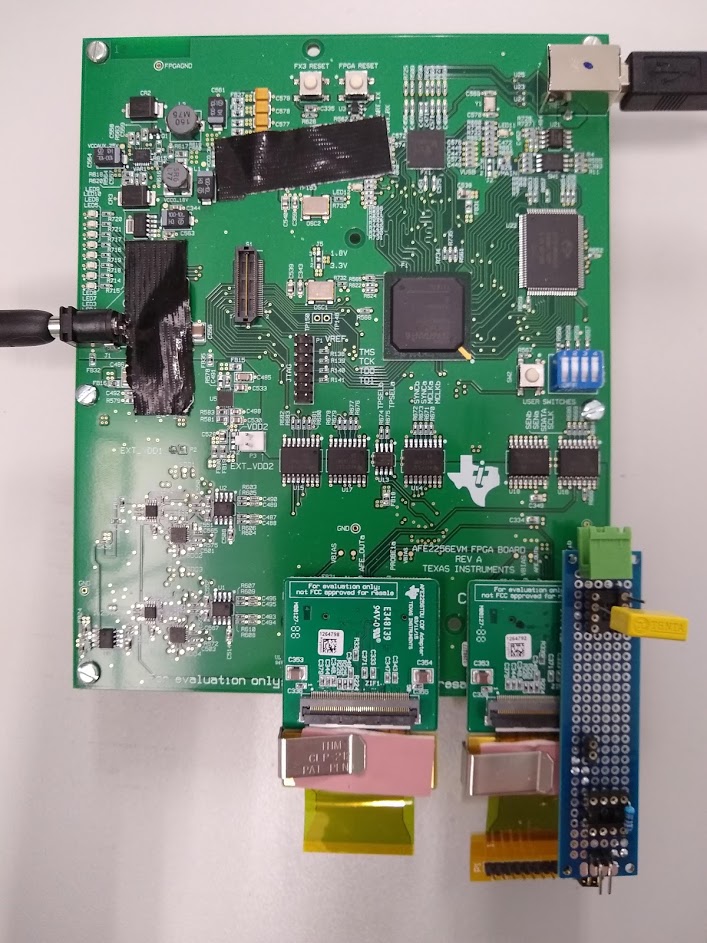}\par
\end{multicols}
\end{center}
\caption{LEFT: Three out of four FMC168 FPGA Mezzanine Cards from Abaco Systems connected to a HTG-840 evaluation platform from Xilinx. This readout system will be operated in pulse coincidence mode. LEFT: The AFE2256EVM analog front end evaluation module from Texas Instruments. This readout system will be operated in charge integration mode.}
\label{fig:electronics}
\end{figure}

We are currently developing two readout systems to be capable of observing a wide range of sub-second signals. The first readout system consists of the 32-channel HTG-840 development platform with the VU440 field programmable gate array (FPGA) from Xilinx and FMC168 FPGA Mezzanine Card from Abaco systems (Fig. \ref{fig:electronics}). Each FMC168 Mezzanine Card has eight 16-bit analog to digital converter (ADC) channels with a sampling rates of 250 Msps per channel allowing us to sample variations down to 8 ns. Each detector channel will be digitized using its own ADC channel (Fig. \ref{fig:HTG840}). We will use a 40 Gbps Peripheral Component Interconnect Express (PCIe) to transfer data from the FPGA to a computer \cite{electme,electno}. The data processing flow can be seen in the left side of Fig. \ref{fig:DataProcess}. This readout system will be operated in the pulse coincidence mode described in section \ref{sec:pulsecoincidence}.

\begin{figure} [H]
\begin{center}
\includegraphics[height=7cm]{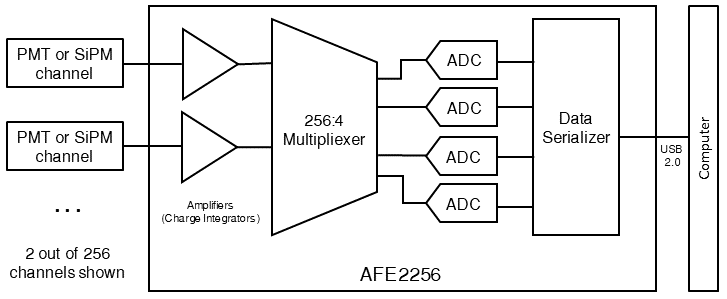}
\end{center}
\caption[example] 
{Schematic of the readout system using the AFE2256EVM evaluation module from Texas Instruments.}
\label{fig:AFE}
\end{figure}

The second readout system consists of the 256-channel AFE2256EVM analog front end evaluation module from Texas Instruments (Fig. \ref{fig:electronics}). The AFE2256EVM is an evaluation platform for AFE2256 chip-on-flex (COF) devices and was originally designed to read out digital X-ray flat panel detectors but has the capability of reading out other detectors such as SiPM as well. We will use a USB 2.0 to transfer data from the AFE2256EVM to a computer. The AFE2256EVM multiplexes using four 16-bit ADCs (Fig. \ref{fig:AFE}) and has a minimum sampling period of 12.8 $\mu$s. The data processing flow can be seen in the right side of Fig. \ref{fig:DataProcess}. The AFE2256EVM will be operated in the charge integration mode described in section \ref{sec:chargeint}.

\subsubsection{Pulse Coincidence Mode}
\label{sec:pulsecoincidence}

Pulse coincidence mode is advantageous for detecting fast and bright transients with pulse durations on the order of the recovery time of the detector. Pulses produced by PMTs and SiPMs have heights proportional to the number of photons detected. As high photon-flux events will produce pulses that are higher than low photon-flux events, a threshold can be placed to filter out sky background. To decrease the number of false alarms caused by random internal noise, we operate two detectors in coincidence and set the coincidence window equal to the recovery interval. A signal that produces a pulse that exceeds a predetermined threshold and is detected by both detectors within the coincidence interval will be recorded as a candidate signal. The event will be time stamped, saved in a central computer for further analysis, and compared with other multi-messenger observatories to investigate its correlation with our target FRB. After a candidate signal is recorded, we will continue observation to look for repetition. 

\subsubsection{Charge Integration Mode}
\label{sec:chargeint}

Charge integration mode is advantageous for detecting weak signals with pulse durations much longer than the recovery time of the detector. The total charge output by the detector in a given time interval will be recorded and averaged to filter out light from the sky background. We look for significant increases in charge within each time interval and repetition. A candidate signal will be time stamped and transmitted to a central computer where it will be saved for further examination and compared with other multi-messenger observatories.

\section{Conclusion}

We are developing a first generation fast, single-photon resolution observatory capable of characterizing the optical (320 nm - 650 nm) sky from the millisecond to nanosecond timescales. Searching in this largely unexplored phase space opens up the possibility of discovering new astrophysical phenomena. While one of the first objectives of the UFA observatory will be to search for and further constrain the limits to optical counterparts to FRB, the UFA observatory will have the capability of discovering a wide range of astrophysical transients. 

We are currently in the preliminary selection and design phase of the project. After performing detector characterization studies and selecting an optimal detector, we will build a prototype camera with two detector arrays and test the effectiveness of the coincidence scheme at lowering the false alarm rate. We will then commission the camera on the NUTTelA-TAO to test and refine our system.

After completion of the first generation UFA observatory, we plan to upgrade our system with 1) larger detector arrays to give us higher spatial resolution, 2) a faster readout system that will increase our signal to noise ratio and allow us to explore faster timescales and 3) different detectors that will allow us to expand our search for counterparts to FRB to longer wavelengths.

\acknowledgments 
 
This work has been supported by the Hong Kong University of Science and Technology, Energetic Cosmos Laboratory at Nazarbayev University, and RK MES grant AP05135753. Siyang would also like to thank those at Nazarbayev University for their hospitality.  

\bibliography{report} 

\begin{thebibliography}{10}

\bibitem{Lorimer777}
Lorimer, D.~R., Bailes, M., McLaughlin, M.~A., Narkevic, D.~J., and Crawford,
  F., ``A {B}right {M}illisecond {R}adio {B}urst of {E}xtragalactic {O}rigin,''
  {\em Science}~{\bf 318},  777--780 (Nov. 2007).

\bibitem{Horowitz2001}
Horowitz, P., M.~Coldwell, C., B.~Howard, A., W.~Latham, D., Stefanik, R.,
  Wolff, J., and M.~Zajac, J., ``{Targeted and all-sky search for nanosecond
  optical pulses at Harvard-Smithsonian},'' in [{\em The Search for
  Extraterrestrial Intelligence (SETI) in the Optical Spectrum
  III}{\nolinebreak\hspace{0.1em}]},  {\em Proc. SPIE}~{\bf 4273},  119--127
  (Aug. 2001).

\bibitem{Eikenberry_1997}
Eikenberry, S.~S., Fazio, G.~G., Ransom, S.~M., Middleditch, J., Kristian, J.,
  and Pennypacker, C.~R., ``{High Time Resolution Infrared Observations of the
  Crab Nebula Pulsar and the Pulsar Emission Mechanism},'' {\em The
  Astrophysical Journal}~{\bf 477},  465--474 (Mar. 1997).

\bibitem{Leung:2018}
Leung, C., Hu, B., Harris, S., Brown, A., Nguyen, H., and Gallicchio, J.,
  ``{Testing the Weak Equivalence Principle using Optical and Near-Infrared
  Crab Pulses},'' {\em The Astrophysical Journal}~{\bf 861},  66 (July 2018).

\bibitem{Petroff2016}
{Petroff}, E., {Barr}, E.~D., {Jameson}, A., {Keane}, E.~F., {Bailes}, M.,
  {Kramer}, M., {Morello}, V., {Tabbara}, D., and {van Straten}, W., ``{FRBCAT:
  The Fast Radio Burst Catalogue},'' {\em {}Publications of the Astronomical
  Society of Australia} .

\bibitem{Amiri2019}
Amiri, M., Bandura, K., Bhardwaj, M., Boubel, P., Boyce, M.~M., Boyle, P.~J.,
  Brar, C., Burhanpurkar, M., Chawla, P., Cliche, J.~F., Cubranic, D., Deng,
  M., Denman, N., Dobbs, M., Fandino, M., Fonseca, E., Gaensler, B.~M.,
  Gilbert, A.~J., Giri, U., Good, D.~C., Halpern, M., Hanna, D., Hill, A.~S.,
  Hinshaw, G., H{\"o}fer, C., Josephy, A., Kaspi, V.~M., Landecker, T.~L.,
  Lang, D.~A., Masui, K.~W., Mckinven, R., Mena-Parra, J., Merryfield, M.,
  Milutinovic, N., Moatti, C., Naidu, A., Newburgh, L.~B., Ng, C., Patel, C.,
  Pen, U., Pinsonneault-Marotte, T., Pleunis, Z., Rafiei-Ravandi, M., Ransom,
  S.~M., Renard, A., Scholz, P., Shaw, J.~R., Siegel, S.~R., Smith, K.~M.,
  Stairs, I.~H., Tendulkar, S.~P., Tretyakov, I., Vanderlinde, K., Yadav, P.,
  and Collaboration, T.~C., ``Observations of fast radio bursts at frequencies
  down to 400 megahertz,'' {\em Nature}~{\bf 566},  230--234 (Jan. 2019).

\bibitem{Gajjar2018}
{Gajjar}, V., {Siemion}, A.~P.~V., {Price}, D.~C., {Law}, C.~J., {Michilli},
  D., {Hessels}, J.~W.~T., {Chatterjee}, S., {Archibald}, A.~M., {Bower},
  G.~C., {Brinkman}, C., {Burke-Spolaor}, S., {Cordes}, J.~M., {Croft}, S.,
  {Enriquez}, J.~E., {Foster}, G., {Gizani}, N., {Hellbourg}, G., {Isaacson},
  H., {Kaspi}, V.~M., {Lazio}, T.~J.~W., {Lebofsky}, M., {Lynch}, R.~S.,
  {MacMahon}, D., {McLaughlin}, M.~A., {Ransom}, S.~M., {Scholz}, P.,
  {Seymour}, A., {Spitler}, L.~G., {Tendulkar}, S.~P., {Werthimer}, D., and
  {Zhang}, Y.~G., ``{Highest Frequency Detection of FRB 121102 at 4-8 GHz Using
  the Breakthrough Listen Digital Backend at the Green Bank Telescope},'' {\em
  The Astrophysical Journal Letters}~{\bf 863},  2 (Aug. 2018).

\bibitem{Kashiyama_2013}
Kashiyama, K., Ioka, K., and M{\'{e}}sz{\'{a}}ros, P., ``Cosmological {F}ast
  {R}adio {B}ursts from {B}inary {W}hite {D}warf {M}ergers,'' {\em The
  Astrophysical Journal}~{\bf 776},  L39 (Oct. 2013).

\bibitem{Geng_2015}
Geng, J.~J. and Huang, Y.~F., ``Fast {R}adio {B}ursts: Collisions between
  {N}eutron {S}tars and {A}steroids/{C}omets,'' {\em The Astrophysical
  Journal}~{\bf 809},  24 (Aug. 2015).

\bibitem{10.1093/mnras/stv2948}
Cordes, J.~M. and Wasserman, I., ``{Supergiant {P}ulses from {E}xtragalactic
  {N}eutron {S}tars},'' {\em Monthly Notices of the Royal Astronomical
  Society}~{\bf 457},  232--257 (Mar. 2016).

\bibitem{Thornton2013}
Thornton, D., Stappers, B., Bailes, M., Barsdell, B., Bates, S., D~R~Bhat, N.,
  Burgay, M., Burke-Spolaor, S., J~Champion, D., Coster, P., D'Amico, N.,
  Jameson, A., Johnston, S., Keith, M.-M., Kramer, M., Levin, L., Milia, S.,
  Ng, C., Possenti, A., and Van~Straten, W., ``A population of fast radio
  bursts at cosmological distances,'' {\em Science}~{\bf 341},  53--56 (July
  2013).

\bibitem{Pen_2015}
Pen, U. and Connor, L., ``{Local Circumnuclear Magnetar Solution to
  Extragalactic Fast Radio Bursts},'' {\em The Astrophysical Journal}~{\bf
  807},  179 (July 2015).

\bibitem{Falcke2013}
Falcke, H. and Rezzolla, L., ``Fast radio bursts: The last sign of supramassive
  neutron stars,'' {\em Astronomy and Astrophysics}~{\bf 562} (Feb. 2014).

\bibitem{Conner2016}
{Connor}, L., {Sievers}, J., and {Pen}, U.-L., ``{{Non-Cosmological FRBs from
  Young Supernova Remnant Pulsars}},'' {\em Monthly Notices of the Royal
  Astronomical Society}~{\bf 458},  L19--L23 (May 2015).

\bibitem{Piro:2016aac}
Piro, A.~L., ``{The Impact of a Supernova Remnant on Fast Radio Bursts},'' {\em
  The Astrophysical Journal}~{\bf 824},  L32 (June 2016).

\bibitem{Kumar2017}
{Kumar}, P., {Lu}, W., and {Bhattacharya}, M., ``{Fast radio burst source
  properties and curvature radiation model},'' {\em Monthly Notices of the
  Royal Astronomical Society}~{\bf 468},  2726--2739 (Jul. 2017).

\bibitem{Spitler2016}
{Spitler}, L.~G., {Scholz}, P., {Hessels}, J.~W.~T., {Bogdanov}, S., {Brazier},
  A., {Camilo}, F., {Chatterjee}, S., {Cordes}, J.~M., {Crawford}, F.,
  {Deneva}, J., {Ferdman}, R.~D., {Freire}, P.~C.~C., {Kaspi}, V.~M.,
  {Lazarus}, P., {Lynch}, R., {Madsen}, E.~C., {McLaughlin}, M.~A., {Patel},
  C., {Ransom}, S.~M., {Seymour}, A., {Stairs}, I.~H., {Stappers}, B.~W., {van
  Leeuwen}, J., and {Zhu}, W.~W., ``{A repeating fast radio burst},'' {\em
  Nature}~{\bf 531},  202--205 (Mar. 2016).

\bibitem{Amiri2019b}
Amiri, M., Bandura, K., Bhardwaj, M., Boubel, P., M~Boyce, M., J~Boyle, P.,
  Brar, C., Burhanpurkar, M., Cassanelli, T., Chawla, P., Cliche, J.-F.,
  Cubranic, D., Deng, M., Denman, N., Dobbs, M., Fandino, M., Fonseca, E.,
  M~Gaensler, B., J~Gilbert, A., and Author, C., ``A second source of repeating
  fast radio bursts,'' {\em Nature}~{\bf 566},  235–238 (Feb. 2019).

\bibitem{Hardy2017}
Hardy, L., Dhillon, V., Spitler, L., Littlefair, S., Ashley, R., De~Cia, A.,
  Green, M., Jaroenjittichai, P., Keane, E., Kerry, P., Kramer, M., Malesani,
  D., Marsh, T., Parsons, S., Possenti, A., Rattanasoon, S., and Sahman, D.,
  ``{A search for optical bursts from the repeating fast radio burst FRB
  121102},'' {\em Monthly Notices of the Royal Astronomical Society}~{\bf 472}
  (Aug. 2017).

\bibitem{Grossan2017}
{Grossan}, B., {Kistaubayev}, M., {Smoot}, G., and {Scherr}, L., ``{Measurement
  of the Shape of the Optical-IR Spectrum of Prompt Emission from Gamma-Ray
  Bursts},'' in [{\em American Astronomical Society Meeting Abstracts
  \#230}{\nolinebreak\hspace{0.1em}]},  {\em American Astronomical Society
  Meeting Abstracts} {\bf 230},  314.06 (June 2017).

\bibitem{Grossan2019}
Grossan, B., Kumar, P., and Smoot, G., ``{The Emission Mechanism of Gamma-ray
  Bursts: Identification via Optical-IR Slope Measurements},'' {\em Journal of
  High Energy Astrophysics}  (2019).
\newblock Manuscript submitted for publication.

\bibitem{datasheet}
{Hamamtsu Photonics K.K.}, ``{S13360 Series MPPC Datasheet}.''
  \url{https://www.hamamatsu.com/resources/pdf/ssd/s13360_series_kapd1052e.pdf}
  (2016).

\bibitem{datasheetb}
{Hamamtsu Photonics K.K.}, ``{S14160/S14161 Series MPPC Datasheet}.''
  \url{https://www.hamamatsu.com/resources/pdf/ssd/s14160_s14161_series_kapd1064e.pdf}
  (2019).

\bibitem{datasheetPMT}
{Hamamtsu Photonics K.K.}, ``{H9500, H9500-03 PMT Datasheet}.''
  \url{https://www.hamamatsu.com/resources/pdf/etd/H9500_H9500-03_TPMH1309E.pdf}
  (2015).

\bibitem{Adams2015}
Adams, J.~H. and et~al., ``The {JEM-EUSO} mission: An introduction,'' {\em
  Experimental Astronomy}~{\bf 40},  3--17 (Nov. 2015).

\bibitem{Atwood:2009ez}
Atwood, W. B. e.~a., ``{The Large Area Telescope on the Fermi Gamma-ray Space
  Telescope Mission},'' {\em The Astrophysical Journal}~{\bf 697},  1071--1102
  (2009).

\bibitem{HOWARD200778}
Howard, A., Horowitz, P., Mead, C., Sreetharan, P., Gallicchio, J., Howard, S.,
  Coldwell, C., Zajac, J., and Sliski, A., ``{Initial results from Harvard
  all-sky optical SETI},'' {\em Acta Astronautica}~{\bf 61},  78 -- 87 (Oct.
  2007).
\newblock Selected Proceedings of the 57th IAF Congress, Valencia, Spain, 2-6
  October, 2006.

\bibitem{Teshima2007}
Teshima, M., Dolgoshein, B., Mirzoyan, R., Nincovic, J., and Popova, E.,
  ``{SiPM Development for Astroparticle Physics Applications},'' {\em
  Proceedings of the 30th International Cosmic Ray Conference, ICRC 2007}~{\bf
  5} (Oct. 2007).

\bibitem{Guberman2017}
Guberman, D., Cortina, J., E.~Ward, J., Hahn, A., Mazin, D., Boix, J.,
  Dettlaff, A., Fink, D., Gaweda, J., Haberer, W., Illa, J.~m., Mundet, J.,
  Vera, Y., and Wetteskind, H., ``{Light-Trap: A SiPM Upgrade for Very High
  Energy Astronomy and Beyond},'' {\em Nuclear Instruments and Methods in
  Physics Research Section A: Accelerators, Spectrometers, Detectors and
  Associated Equipment}~{\bf 912} (Sept. 2017).

\bibitem{Wright2018}
Wright, S., Horowitz, P., Maire, J., Werthimer, D., Antonio, F., Aronson, M.,
  Chaim-Weismann, S.~A., Cosens, M., Drake, F.~D., W.~Howard, A., Marcy, G.~W.,
  Raffanti, R., Siemion, A. P.~V., Stone, R. P.~S., Treffers, R.~R., and
  Uttamchandani, A., ``Panoramic optical and near-infrared {SETI} instrument:
  overall specifications and science program,'' in [{\em Ground-based and
  Airborne Instrumentation for Astronomy VII}{\nolinebreak\hspace{0.1em}]},
  {\em Proc. SPIE} {\bf 10702},  107025I (July 2018).

\bibitem{Li2019}
Li, S., Maire, J., Cosens, M., and Wright, S.~A., ``Detector characterization
  of a near-infrared discrete avalanche photodiode 5x5 array for astrophysical
  observations,'' in [{\em Infrared Technology and Applications
  XLV}{\nolinebreak\hspace{0.1em}]},  {\em Proc. SPIE} {\bf 11002},  110022G
  (May 2019).

\bibitem{Li2019b}
Li, S. and Smoot, G.~F., ``{Characterization of a silicon photomultiplier for
  the Ultra-Fast Astronomy telescope},'' {\em Proc. SPIE}  (2019).
\newblock Manuscript submitted for publication.

\bibitem{electme}
Shafiee, M., Feghhi, S. A.~H., and Rahighi, J., ``{Experimental performance
  evaluation of ILSF BPM data acquisition system},'' {\em Measurement}~{\bf
  100},  205--212 (March 2017).

\bibitem{electno}
Shafiee, M., Feghhi, S. A.~H., and Rahighi, J., ``{Analysis of de-noising
  methods to improve the precision of the ILSF BPM electronic readout
  system},'' {\em Journal of Instrumentation}~{\bf 11} (December 2016).

\end{thebibliography}
\bibliographystyle{spiebib} 

\end{document}